\documentclass[11pt,draftclsnofoot,onecolumn]{IEEEtran}

\IEEEoverridecommandlockouts

\usepackage{url}
\usepackage{ifthen}
\usepackage[cmex10]{amsmath}
\usepackage{amssymb}
\usepackage{stackrel}

\usepackage{graphicx}
\usepackage{cite}
\usepackage{enumerate}

\usepackage[caption=false,font=footnotesize]{subfig}  

\newtheorem{defi}{Definition}
\newtheorem{theorem}{Theorem}

\newtheorem{remark}{Remark}

\newcommand{\Xc}{\mathcal{X}}
\newcommand{\Yc}{\mathcal{Y}}

\newcommand{\Cc}{\mathcal{C}}
\newcommand{\Mc}{\mathcal{M}}

\newcommand{\Rc}{\mathcal{R}}

\newcommand{\Nc}{\mathcal{N}}

\begin{document}
%
\title{Constrained Colluding Eavesdroppers: An Information-Theoretic Model}

%
%

\author{Mahtab Mirmohseni and Panagiotis Papadimitratos\\
KTH Royal Institute of Technology, Stockholm, Sweden \\
Email: \{mahtabmi,papadim\}@kth.se}

\maketitle

\begin{abstract}
We study the secrecy capacity in the vicinity of colluding eavesdroppers. Contrary to the \emph{perfect collusion} assumption in previous works, our new information-theoretic model considers \emph{constraints} in collusion. We derive the achievable secure rates (lower bounds on the perfect secrecy capacity), both for the discrete memoryless and Gaussian channels. We also compare the proposed rates to the non-colluding and perfect colluding cases.
\end{abstract}


\IEEEpeerreviewmaketitle


\section{Introduction}
Wyner \cite{Wyn75} introduced the information-theoretic model for confidentiality in noisy communications, called \emph{wiretap channel}, where a legitimate transmitter wishes to transmit a confidential message to a legitimate receiver while keeping it hidden from an eavesdropper (wiretapper). The eavesdropper is assumed to have unlimited computation power, know the coding scheme of the legitimate user, and it only listens to the channel. When the channel to the eavesdropper is a degraded version of the channel to the legitimate receiver, Wyner \cite{Wyn75} proposed the secrecy capacity achieving scheme, known also as \emph{Wyner's wiretap channel coding}, which constitutes of multicoding and randomized encoding \cite[Section~22.1.1]{ElgKim11}. This result is extended to the broadcast channel with confidential message and to the general wiretap channel (not necessarily degraded) by Csisz\'{a}r and K\"{o}rner \cite{CsiKor78}.

Recently, different legitimate-wiretapper user combinations were studied~\cite{LiuMarSpaYat08,EkrUlu13,ChiaElG12,Ooh07,LaiElG08}. In this line of works, scenarios with multiple eavesdroppers considered only \emph{non-colluding} ones. This implies that information leakage of a certain message to all eavesdropper is computed as the maximum of the leakage to each of them. In some applications, this assumption may underestimate the eavesdroppers' power: eavesdroppers can collude, i.e., share their channel outputs (observations), and make the attack more effective \cite{PinBarWin09}. Hence, combating colluding eavesdroppers, especially in wireless networks, has been a significant challenge~\cite{PinBarWin09,KoyKokElg12,ZhaFuWan12,PinBarWin12II,GoeNeg05,WanHuaWan13}. To the best of our knowledge, all previous works modeled $k$ colluding eavesdroppers as one eavesdropper with $k$ antennas; we term this \emph{perfect colluding} eavesdroppers. Using the equivalent Single-Input Multiple-Output (SIMO) Gaussian wiretap channel, the information leakage is determined by the aggregate Signal to Noise Ratio (SNR) of all eavesdroppers; compared to the maximum SNR in the non-colluding case \cite{PinBarWin09}. This assumption significantly overestimates eavesdropping capability, forcing a legitimate user to increase its power linearly with the number of eavesdroppers to achieve a positive secure rate. However, collusion (esp. in the wireless networks) necessitates communication resources and power consumption. This, in fact, restricts the collusion channel capacity and thus improves the achievable secure rate by the legitimate user. Hence, here the problem is to find an appropriate model and to analyze the effect of these constraints on the secrecy capacity based on this model.

\subsection{Our Contributions}\label{subsec:contributions}
In this paper, we consider the potential constraints in collusion, by modeling \emph{constrained collusion} with an equivalent wiretap channel, called \emph{Wiretap Channel with Constrained Colluding Eavesdroppers} (WTC-CCE). For our \emph{general} WTC-CCE, we assume that colluding eavesdroppers communicate (by defining their channel inputs) over a virtual \emph{collusion channel}, in addition to the main channel. The higher the collusion channel capacity, the more leaked information can be exchanged. Our model captures previously studied models as special cases: non-colluding with zero collusion rates and perfect collusion with infinite collusion rates. We also propose a special case, the \emph{orthogonal} WTC-CCE: the collusion channel is orthogonal to the main one (unlike the general WTC-CCE where eavesdroppers shares the same channel with the legitimate transmitter). First, we derive an achievable secure rate (a lower bound on the perfect secrecy capacity) for the general discrete memoryless WTC-CCE. The idea is to let the eavesdroppers do their best in colluding. Hence, the information leakage rate is derived by considering the outer bound on the capacity region of the collusion channel; this resembles the cut-set upper bound for the relay channel \cite{ElgKim11}. Next, we extend our result to the general Gaussian WTC-CCE and its orthogonal version. The main difference is that in the general model, the eavesdroppers may use jamming techniques to confuse the legitimate receiver but they could be exposed to the legitimate user. In the orthogonal model, beyond increased required resources, the eavesdroppers may loose some information leakage rate due to not sending jamming signals. However, the orthogonality may serve eavesdroppers in hiding themselves. We provide  numerical examples to analyze the achievable secure rate and evaluate the overestimation amount (by comparing to perfect colluding case) in different scenarios.

The rest of the paper is organized as follows. Section~\ref{sec:definition} introduces the channel model and the notations. In Section~\ref{sec:DMC}, our main results for the general discrete-memoryless channel are presented, while in Section~\ref{sec:Gaus}, the Gaussian channel results are stated. Finally, Section~\ref{sec:discussion} concludes the paper.

\section{Channel Model and Preliminaries}\label{sec:definition}
\begin{figure}[tb]
  \centering
  \includegraphics[width=9cm]{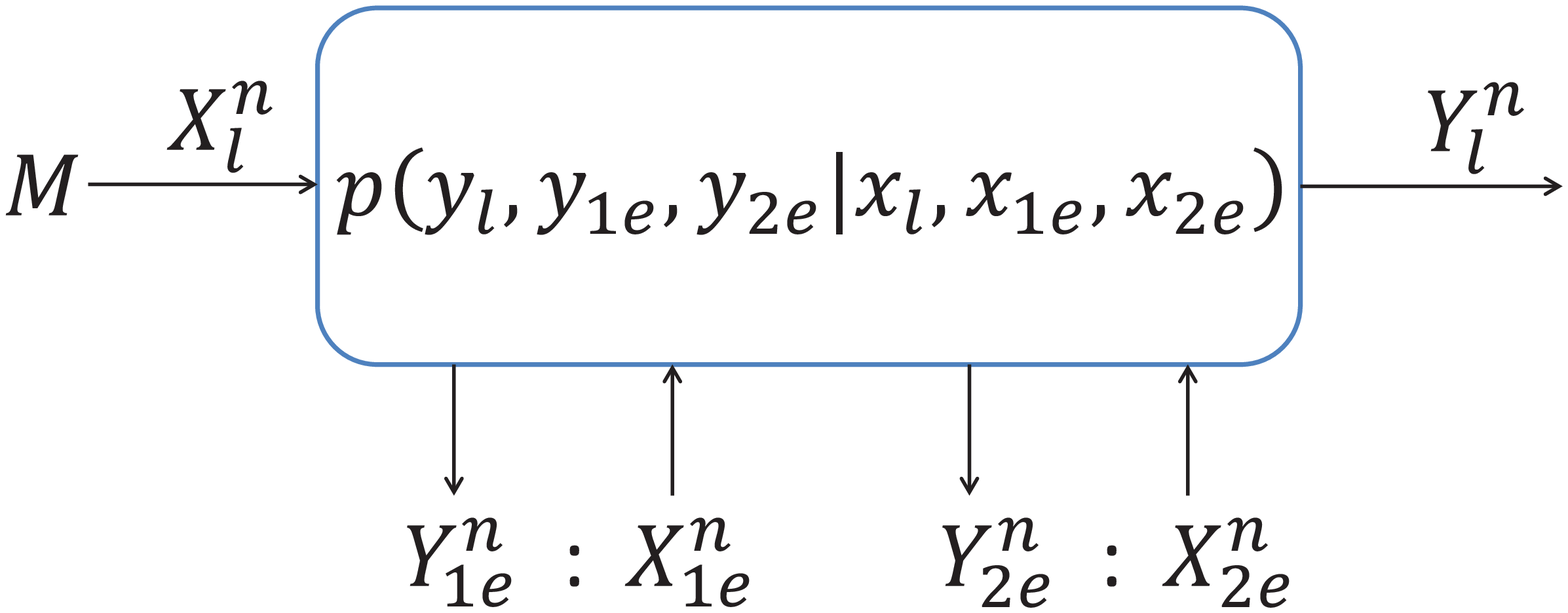}
  \caption{General Wiretap Channel with Constrained Colluding Eavesdroppers (WTC-CCE).}
  \label{fig:CE_BD_gen}
\end{figure}
Upper-case letters (e.g., $X$) denote Random Variables (RVs) and lower-case letters (e.g., $x$) their realizations. The probability mass function (p.m.f) of a RV $X$ with alphabet set $\Xc$ is denoted by $p_X(x)$; occasionally, the subscript $X$ is omitted. $X^j_i$ indicates a sequence of RVs $(X_i,X_{i+1},...,X_j)$; we use $X^j$ instead of $X^j_1$ for brevity. $\Nc(0,\sigma^2)$ denotes a zero-mean Gaussian distribution with variance $\sigma^2$.

Consider the WTC-CCE in Fig.~\ref{fig:CE_BD_gen}: a four terminal discrete channel (one transmitter, one legitimate receiver and two eavesdroppers), denoted by ($\Xc_l\times\Xc_{1e}\times\Xc_{2e},p(y_l^n,y_{1e}^n,y_{2e}^n|x_l^n,x_{1e}^n,x_{2e}^n),\Yc_l\times\Yc_{1e}\times\Yc_{1e}$). $X_l\in\Xc_l$ and $X_{je}\in\Xc_{je}$ are the channel inputs of the legitimate transmitter and eavesdropper~$j$ and $Y_l\in\Yc_l$ and $Y_{je}\in\Yc_{je}$ are the channel outputs at the legitimate receiver and eavesdropper~$j$, for $j\in\{1,2\}$. $p(y_l^n,y_{1e}^n,y_{2e}^n|x_l^n,x_{1e}^n,x_{2e}^n)$ is the channel transition probability distribution. We also assume that the channel is memoryless. In $n$ channel uses, the legitimate transmitter desires to send the message $M$ to the legitimate receiver using the following code.
\begin{defi}\label{def:code}
A $(2^{nR},n,P_e^{(n)})$ code for WTC-CCE consists of:
 \begin{enumerate}[(i)]
   \item A message set $\Mc=[1:2^{nR}]$, where $m$ is uniformly distributed over $\Mc$.
   \item A \emph{randomized} encoding function, $f_{n}$, at the legitimate transmitter that maps a message $m$ to a codeword $x_l^n\in\Xc_l^n$.
   \item Two sets of encoding functions at the eavesdroppers: $\{f_{je,t}\}_{t=1}^{n}:\mathbb{R}^{t-1}\longrightarrow \mathbb{R}$ such that $x_{je,t}=f_{je,t}(y_{je}^{t-1})$, for $j\in\{1,2\}$ and $1\leq t\leq n$.
   \item A decoding function at the legitimate receiver $g: \Yc_{l}^{n}\mapsto\Mc$.
   \item Probability of error for this code is defined as: 
   \begin{align}\label{eqn:def_Pe}
      P_e^{(n)}=\frac{1}{2^{nR}}\sum\limits_{m\in\Mc}{Pr(g(y_{l}^{n})\neq m | m\textrm{ sent})}.
   \end{align}
   \item The information leakage rate at eavesdropper $j\in\{1,2\}$ is defined as:
    \begin{align}\label{eqn:def_leakage}
        R_{L,j}^{(n)}=\frac{1}{n}I(M;Y_{je}^n).
    \end{align}
 \end{enumerate}

All codewords are revealed to the eavesdroppers. However, eavesdroppers' mapping are not known to the legitimate user.
\end{defi}
\begin{remark}\label{remark:model}
The mutual information term in \eqref{eqn:def_leakage} is same as the non-colluding case, compared to $I(M;Y_{1e}^n,Y_{2e}^n)$ in the perfect colluding scenario. The difference here comes from the channel distribution and the fact that $Y_{1e}^n$ and $Y_{1e}^n$ given $X_l$ are not independent (due to $X_{1e}$ and $X_{2e}$).
\end{remark}
\begin{defi}\label{def:rate}
A rate-leakage tuple $(R,R_{L,1},R_{L,2})$ is achievable if there exists a sequence of $(2^{nR},n,P_e^{(n)})$ codes such that $P_e^{(n)}\rightarrow 0$ as $n\rightarrow\infty$ and $\limsup\limits_{n\rightarrow\infty} R_{L,j}^{(n)}\leq R_{L,j}$ for $j\in\{1,2\}$. The secrecy capacity $\Cc_s$ is the supremum of all achievable rates $R$ such that perfect secrecy is achieved, i.e., $R_{L,j}=0$ for $j\in\{1,2\}$.
\end{defi}

\begin{figure}[tb]
  \centering
  \includegraphics[width=9cm]{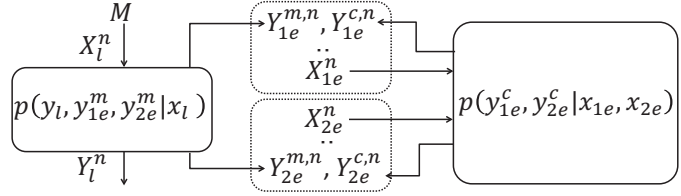}
  \caption{Orthogonal WTC-CCE.}
  \label{fig:CE_BD_orth}
\end{figure}
Motivated by the fact that the eavesdroppers prefer to avoid exposure, we also consider a special case of the WTC-CCE. We assume that the collusion channel (used by the eavesdroppers) is decoupled from the main channel and consider the orthogonal WTC-CCE in Fig.~\ref{fig:CE_BD_orth}. Here, $Y_{je}=(Y_{je}^m,Y_{je}^c)$ for $j\in\{1,2\}$ and $p(y_l,y_{1e},y_{2e}|x_l,x_{1e},x_{2e})=p(y_l,y_{1e}^m,y_{2e}^m|x_l)p(y_{1e}^c,y_{2e}^c|x_{1e},$ $x_{2e})$, where the variables related to the main and the collusion channels are indicated with the superscripts $m$ and $c$, respectively. Substituting $X_{1e}=X_{2e}=\emptyset$ results in the non-colluding case; $Y_{1e}^c=Y_{2e}^m , Y_{2e}^c=Y_{1e}^m$ results in the perfect colluding case.
To simplify notation let $\bar{j}$ be the complement of $j$ in $\{1,2\}$. Now, consider the general Gaussian WTC-CCE at time $t=1,\ldots,n$ for $j\in\{1,2\}$, modeled as:
\begin{IEEEeqnarray}{rcl}
Y_{l,t}&\:=\:&h_{l}X_{l,t}+h_{1e}^lX_{1e,t}+h_{2e}^lX_{2e,t}+Z_{l,t}\nonumber\\
Y_{je,t}&\:=\:&h_{l}^{je}X_{l,t}+h_{\bar{j}e}^{je}X_{\bar{j}e,t}+Z_{je,t}\label{eqn:Gen_Gaussian_model}
\end{IEEEeqnarray}
where $h_{i}^k$ is a known channel gain from transmitter $i$ to receiver $k$. We assume perfect echo cancellation at eavesdroppers ($h_{1e}^{1e}=h_{2e}^{2e}=0$). $X_{u,t}$ is an input signal with average power constraint
\begin{IEEEeqnarray}{rcl}
\frac{1}{n}\sum\limits_{t=1}^n|x_{u,t}|^2\leq P_u\label{eqn:power_cons}
\end{IEEEeqnarray}
and $Z_{u,t}$ is an independent and identically distributed (i.i.d) zero-mean Gaussian noise component with power $N_u$, for $u\in\{l,1e,2e\}$. In practice, $h_{1e}^l$ and $h_{2e}^l$ may be small. The Gaussian counterpart of orthogonal WTC-CCE for $j\in\{1,2\}$ can be shown as:
\begin{IEEEeqnarray*}{rcl}
Y_{l,t}&\:=\:&h_{l}X_{l,t}+Z_{l,t}\yesnumber\label{eqn:Orth_Gaussian_model}\\
Y_{je,t}^m&\:=\:&h_{jm}X_{l,t}+Z_{je,t}^m  \quad,\quad  Y_{je,t}^c\:=\:h_{jc}X_{\bar{j}e,t}+Z_{je,t}^c
\end{IEEEeqnarray*}
where $h_{jm}$ and $h_{jc}$ are known channel gains received at eavesdropper $j$ from the main channel and the collusion channel, respectively; power constraints of $P_l,P_{1e},P_{2e}$ apply for input signals; $Z_{je,t}^m$ and $Z_{je,t}^c$ are i.i.d zero-mean Gaussian noise components with powers $N_{je}^m$ and $N_{je}^c$ at eavesdropper $j$ from the main channel and the collusion channel, respectively.
\section{Discrete Memoryless Channel}\label{sec:DMC}
Our first result establishes an achievable secure rate for the general discrete memoryless WTC-CCE.
\begin{theorem}\label{thm:Ach_gen_DM}
For the general discrete memoryless WTC-CCE, the secrecy capacity is lower-bounded by:
\begin{IEEEeqnarray}{rl}
\Rc_s^{DM}=\sup\inf I(X_l;Y_l)-\min\{&I(X_l;Y_{1e},Y_{2e}|X_{1e},X_{2e}),\nonumber\\
&\max \{I(X_l,X_{1e},X_{2e};Y_{1e}),I(X_l,X_{1e},X_{2e};Y_{2e})\}\}\label{eqn:Ach_gen_DM}
\end{IEEEeqnarray}
where the supremum and infimum are taken over all joint p.m.fs of the form $p(x_l|x_{1e},x_{2e})p(y_l,y_{1e},y_{2e}|x_l,$ $x_{1e},x_{2e})$ and $p(x_{1e},x_{2e})$, respectively.
\end{theorem}
\begin{IEEEproof}
The proof is based on the random coding scheme, which uses Wyner wiretap coding at the legitimate user. At the eavesdroppers, the idea is to let them do their best in colluding. Hence, the coding strategy of the eavesdroppers is not determined in the scheme. As a result, the information leakage rate is derived by considering the outer bound on the capacity region of the collusion channel and looks like the cut-set upper bound for the relay channel \cite{ElgKim11}.

\textit{Codebook Generation:} Generate $2^{n(R+R_s)}$ i.i.d $x_l^{n}$ sequences, each with probability $\prod\limits_{t = 1}^{n}p(x_{l,t})$. Index them as $x_l^{n}(m,s)$ where $m\in[1:2^{nR}]$ and $s\in[1:2^{nR_s}]$.

\textit{Encoding:} To send message $m\in[1:2^{nR}]$, the stochastic encoder at the legitimate transmitter uniformly randomly chooses $s$ and transmits $x_l^{n}(m,s)$.

\textit{Decoding:} The decoder at the legitimate receiver wants to correctly recover $m,s$ and seeks a unique message $\tilde{m}$ and some $\tilde{s}$ such that
$(x_l^{n}(\tilde{m},\tilde{s}),y_l^n)$ are jointly typical. Applying the packing lemma \cite{ElgKim11}, with arbitrary high probability $\tilde{m}=m$, if $n$ is large enough and
\begin{IEEEeqnarray}{rcl}
R+R_s&\leq& I(X_l;Y_l). \label{eqn:ach1_I}
\end{IEEEeqnarray}

\textit{Analysis of information leakage rate:} To simplify the notation, let $X_e=(X_{1e},X_{2e})$ and $Y_e=(Y_{1e},Y_{2e})$. We derive two bounds for the randomness index rate, $R_s$. First, we obtain the second term of information leakage rates in the $\min$ term in \eqref{eqn:Ach_gen_DM}, i.e., $R_{L2}=\max \{I(X_l,X_{1e},X_{2e};Y_{1e}),I(X_l,X_{1e},X_{2e};Y_{2e})\}$.

Now, consider the leakaged information to $Y_{1e}^n$ averaged over the random codebook $\Cc$.
\begin{IEEEeqnarray*}{rcl}
I(M;Y_{1e}^n|\Cc)&=&H(M|\Cc)-H(M|Y_{1e}^n,\Cc)\\
&=&nR-H(M,Y_{1e}^n,X_l^n,X_e^n|\Cc)+H(X_l^n,X_e^n|M,Y_{1e}^n,\Cc)+H(Y_{1e}^n|\Cc)\\
&=&nR-H(X_l^n,X_e^n|\Cc)-H(M,Y_{1e}^n|X_l^n,X_e^n,\Cc)+H(X_l^n,X_e^n|M,Y_{1e}^n,\Cc)+H(Y_{1e}^n|\Cc)\\
&{\leq}& nR-H(X_l^n|\Cc)-H(Y_{1e}^n|X_l^n,X_e^n,\Cc)+H(X_l^n,X_e^n|M,Y_{1e}^n,\Cc)+H(Y_{1e}^n|\Cc)\\
&=& nR-n(R+R_s)+I(X_l^n,X_e^n;Y_{1e}^n|\Cc)+H(X_l^n,X_e^n|M,Y_{1e}^n,\Cc)\\
&\stackrel{(a)}{\leq}&-nR_s+nI(X_l,X_e;Y_{1e})+H(X_l^n,X_e^n|M,Y_{1e}^n,\Cc)\stackrel{(b)}{\leq}n\delta_1
\end{IEEEeqnarray*}
(a) holds since the channel is memoryless; (b) follows by using \cite[Lemma~22.1]{ElgKim11}: if $R_s\geq I(X_l,X_{1e},X_{2e};Y_{1e})$, then $H(X_l^n,X_{1e}^n,X_{2e}^n|M,Y_{1e}^n,\Cc)\leq nR_s-nI(X_l,X_{1e},X_{2e};Y_{1e})+n\delta_1$. Following similar steps, one can show that if $R_s\geq I(X_l,X_{1e},X_{2e};Y_{2e})$, then $I(M;Y_{2e}^n|\Cc)\leq \delta_2$. Considering \eqref{eqn:def_leakage}, combining \eqref{eqn:ach1_I} and these constraints on $R_s$ gives $\Rc_s^{DM}$ with $R_{L2}$.

Now, to derive the first term of information leakage rates in $\min$ in \eqref{eqn:Ach_gen_DM}, i.e., $R_{L1}=I(X_l;Y_{1e},Y_{2e}|X_{1e},X_{2e})$, and evaluate the leakaged information to both $Y_{1e}^n$ and $Y_{2e}^n$, averaged over the random codebook $\Cc$.
\begin{IEEEeqnarray*}{rcl}
I(M;Y_e^n|\Cc)&=&H(M|\Cc)-H(M|Y_e^n,\Cc)\\
&=&nR-H(M,Y_e^n,X_l^n|\Cc)+H(X_l^n|M,Y_e^n,\Cc)+H(Y_e^n|\Cc)\\
&\stackrel{(a)}{=}& nR-H(X_l^n|\Cc)-H(M,Y_e^n|X_l^n,\Cc)+H(X_l^n|M,Y_e^n,X_e^n,\Cc)+H(Y_e^n|\Cc)\\
&\stackrel{(b)}{\leq}& nR-n(R+R_s)+I(X_l^n;Y_e^n|\Cc)+H(X_l^n|M,Y_e^n,X_e^n,\Cc)\\
&\stackrel{(c)}{=}&-nR_s+\sum\limits_{i=1}^{n}I(X_l^n;Y_{e,i}|Y_{e}^{i-1},X_{e,i},\Cc)+H(X_l^n|M,Y_e^n,X_e^n,\Cc)\\
&\stackrel{(d)}{\leq}&-nR_s+nI(X_l;Y_e|X_e)+H(X_l^n|M,Y_e^n,X_e^n,\Cc)\stackrel{(e)}{\leq}n\delta_3\yesnumber\label{eqn:ach1_delta3}
\end{IEEEeqnarray*}
(a) and (c) follow since $x_{je,t}=f_{je,t}(y_{je}^{t-1})$, for $j\in\{1,2\}$ and $1\leq t\leq n$; (b) is due to the fact that conditioning does not increase the entropy; (d) holds due to the memoryless property of the channel; (e) follows by using \cite[Lemma~22.1]{ElgKim11}: if $R_s\geq I(X_l;Y_{1e},Y_{2e}|X_{1e},X_{2e})$, then $H(X_l^n|M,Y_{1e}^n,Y_{2e}^n,X_{1e}^n,X_{2e}^n,\Cc)\leq nR_s-nI(X_l;Y_{1e},Y_{2e}|X_{1e},X_{2e})+n\delta_3$. Note that \eqref{eqn:ach1_delta3} implies the individual leakage rates as $I(M;Y_{je}^n|\Cc){\leq}n\delta_3$ for $j\in\{1,2\}$. Now, combining \eqref{eqn:ach1_I} and this contraint on $R_s$ gives $\Rc_s^{DM}$ with $R_{L1}$. This completes the proof.
\end{IEEEproof}
\begin{remark}\label{remark:Ach_gen_DM_orth}
Substituting $Y_{je}=(Y_{je}^m,Y_{je}^c)$ for $j\in\{1,2\}$ in \eqref{eqn:Ach_gen_DM} results in an achievable secure rate ($\Rc_s^{ODM}$) for the orthogonal discrete memoryless WTC-CCE, where the supremum is taken over all joint p.m.fs of the form $p(x_l|x_{1e},x_{2e})p(y_l,y_{1e}^m,y_{2e}^m|x_l)p(y_{1e}^c,y_{2e}^c|x_{1e},x_{2e})$.
\end{remark}
\begin{remark}\label{remark:Ach_gen_DM_nc_pc}
By setting $X_{1e}=X_{2e}=\emptyset$ in \eqref{eqn:Ach_gen_DM}, $\Rc_s^{DM}$ reduces to $\sup I(X_l;Y_l)-\max \{I(X_l;Y_{1e}),I(X_l;Y_{2e})\}$ for the non-colluding case. Furthermore, redefining $Y_{1e}^c=Y_{2e}^m , Y_{2e}^c=Y_{1e}^m$ in $\Rc_s^{ODM}$ results in the achievable secure rate for the perfect colluding case, i.e., $\sup I(X_l;Y_l)- I(X_l;Y_{1e},Y_{2e})$.
\end{remark}
\section{Gaussian Channel}\label{sec:Gaus}
We study the Gaussian WTC-CCE. First, we consider the orthogonal Gaussian WTC-CCE. Let $\theta(x)\doteq \frac{1}{2}\log(1+x)$.
\begin{theorem}\label{thm:Ach_Orth_Gaus}
The following is an achievable secure rate for orthogonal Gaussian WTC-CCE (defined in \eqref{eqn:Orth_Gaussian_model}).
\begin{IEEEeqnarray*}{rl}
\Rc_s^{OG}=\theta(\frac{h_l^2P_l}{N_l})-&\min\Big\{\theta(P_l(\frac{h_{1m}^2}{N_{1e}^m}+\frac{h_{2m}^2}{N_{2e}^m})),\yesnumber\label{eqn:Ach_Orth_Gaus}\\
&\max\{\theta(\frac{h_{1m}^2P_l}{N_{1e}^m}+\frac{h_{1c}^2P_{2e}}{N_{1e}^c}+\frac{h_{1m}^2h_{1c}^2P_lP_{2e}}{N_{1e}^cN_{1e}^m}),
\theta(\frac{h_{2m}^2P_l}{N_{2e}^m}+\frac{h_{2c}^2P_{1e}}{N_{2e}^c}+\frac{h_{2m}^2h_{2c}^2P_lP_{1e}}{N_{2e}^cN_{2e}^m})\}\Big\}.\\
\end{IEEEeqnarray*}
\end{theorem}

\begin{IEEEproof}
We can extend the achievable secrecy rate in Theorem~\ref{thm:Ach_gen_DM} (after applying Remark~\ref{remark:Ach_gen_DM_orth}) to the Gaussian case with continuous alphabets with standard arguments \cite{CovTho06}. As we do not know the optimal distribution $p(x_l|x_{1e},x_{2e})$ that maximizes $\Rc_s^{ODM}$, we use a Gaussian input distribution (at the legitimate transmitter) to achieve a lower bound. Let $X_l\sim\Nc(0,P_l)$. Note that the leakage rates in $\Rc_s^{ODM}$ (i.e., $R_{L1}$ and $R_{L2}$) are Multiple Access Channel (MAC) type bounds. From the maximum-entropy theorem \cite{CovTho06} (or \cite[P.~21]{ElgKim11}), these bounds are largest (or equivalently $\Rc_s^{ODM}$ in minimized over $p(x_{1e},x_{2e})$) for the Gaussian inputs at the eavesdroppers. Hence, set $X_{je}\sim\Nc(0,P_{je})$ for $j\in\{1,2\}$ and define $-1\leq\rho_j\leq 1$ as the correlation coefficient between $X_{je}$ and $X_l$, i.e., $E(X_{je}X_l)=\rho_j\sqrt{P_{je}P_l}$ for $j\in\{1,2\}$ and $\rho_{12}=\frac{E(X_{1e}X_{2e})}{\sqrt{P_{1e}P_{2e}}}$. After, calculating the mutual information terms in \eqref{eqn:Ach_gen_DM}, one can easily show that the leakage rate is maximized (or secure rate in minimized) for $\rho_{12}=\rho_{1}=\rho_{2}=0$. This means that in the orthogonal setup, the best strategy for the eavesdroppers is using the independent codewords. This achieves $\Rc_s^{OG}$ in \eqref{eqn:Ach_Orth_Gaus}.
\end{IEEEproof}
\begin{remark}\label{remark:Ach_Orth_Gaus_nc_pc}
To achieve the non-colluding rate, i.e., $\theta(\frac{h_l^2P_l}{N_l})-\max\{\theta(\frac{h_{1m}^2P_l}{N_{1e}^m}),\theta(\frac{h_{2m}^2P_l}{N_{2e}^m})\}$, set $P_{1e}=P_{2e}=0$ in $\Rc_s^{OG}$. Moreover, it is enough to set $P_{1e},P_{2e}\rightarrow\infty$ in $\Rc_s^{OG}$ to derive the perfect colluding rate: $\theta(\frac{h_l^2P_l}{N_l})-\theta(P_l(\frac{h_{1m}^2}{N_{1e}^m}+\frac{h_{2m}^2}{N_{2e}^m}))$.
\end{remark}
In the following, we obtain a secure rate for the general Gaussian WTC-CCE. The proof is similar to Theorem~\ref{thm:Ach_Orth_Gaus}.
\begin{theorem}\label{thm:Ach_gen_Gaus}
The following is an achievable secure rate for Gaussian WTC-CCE (in \eqref{eqn:Gen_Gaussian_model}).
\begin{IEEEeqnarray*}{rcl}
\Rc_s^{G}&=&\min\limits_{\rho_1,\rho_2,\rho_{12}}\theta(\frac{h_l^2P_l+\rho_1^2(h_{1e}^l)^2P_{1e}+\rho_2^2(h_{2e}^l)^2P_{2e}+2h_lh_{1e}^l\rho_1\sqrt{P_{l}P_{1e}}+2h_lh_{2e}^l\rho_2\sqrt{P_{l}P_{2e}}}{(h_{1e}^l)^2P_{1e}(1-\rho_1^2)+(h_{2e}^l)^2P_{2e}(1-\rho_2^2)+2h_{1e}^lh_{2e}^l\rho_{12}\sqrt{P_{1e}P_{2e}}+N_l})\yesnumber\label{eqn:Ach_Gen_Gaus}\\
&&-\min\Big\{\max\{A(1),A(2)\},
\theta(P_l(1-\frac{\rho_1^2P_{1e}^2+\rho_2^2P_{2e}^2+2\rho_1\rho_2\rho_{12}P_{1e}P_{2e}}{P_{1e}P_{2e}(1-\rho_{12}^2)})(\frac{(h_{l}^{1e})^2}{N_{1e}}+\frac{(h_{l}^{2e})^2}{N_{2e}}))\Big\}.
\end{IEEEeqnarray*}
where for $j\in\{1,2\}$: 
\begin{IEEEeqnarray*}{rcl}
A(j)=\theta\left(\frac{(h_{l}^{je})^2P_l+(h_{\bar{j}e}^{je})^2P_{\bar{j}e}+2h_{l}^{je}h_{\bar{j}e}^{je}\rho_2\sqrt{P_{l}P_{\bar{j}e}}}{N_{je}}\right).
\end{IEEEeqnarray*}
\end{theorem}

\begin{remark}\label{remark:Ach_Gen_Gaus_I}
Channel gains $h_{1e}^l$ and $h_{2e}^l$ make the jamming possible for the eavesdroppers. However, they also increase the probability of exposure. In order to compare the two strategies (through numerical examples), we define the non-jamming rate $\Rc_s^{NJG}$ by setting $h_{1e}^l=h_{2e}^l=0$ in $\Rc_s^{G}$. In addition, by setting $P_{1e},P_{2e}\rightarrow\infty$ in $\Rc_s^{G}$, the secure rate is zero, which is less than (or equal to) the perfect colluding rate. This is due to the jamming possibility and is achieved by $\rho_{12}=\rho_{1}=\rho_{2}=0$.
\end{remark}

\begin{figure*}[!tb]%
    \centering
    \captionsetup[figure]{margin=10pt}%
    \subfloat[$h_{2e}^{1e}=h_{1e}^{2e}=h_{jc}=\sqrt{0.1},j\in\{1,2\}$.\label{fig:Gaus_Case1A}]{\includegraphics[width=7.7cm,height=5cm]{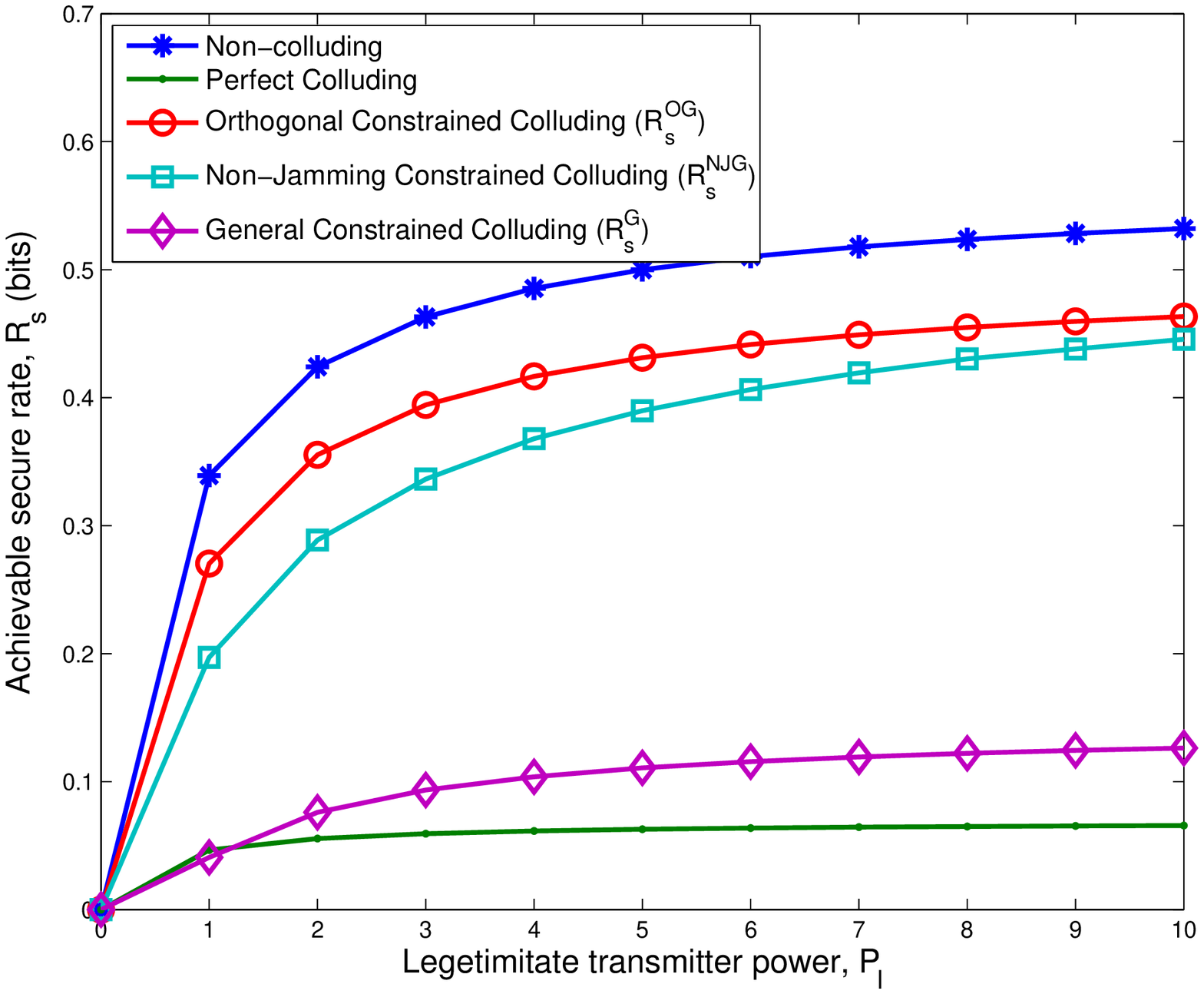}}
    \hspace{20pt}
    \subfloat[$h_{2e}^{1e}=h_{1e}^{2e}=h_{jc}=\sqrt{0.6},j\in\{1,2\}$.\label{fig:Gaus_Case1B}]%
     {\includegraphics[width=7.7cm,height=5cm]{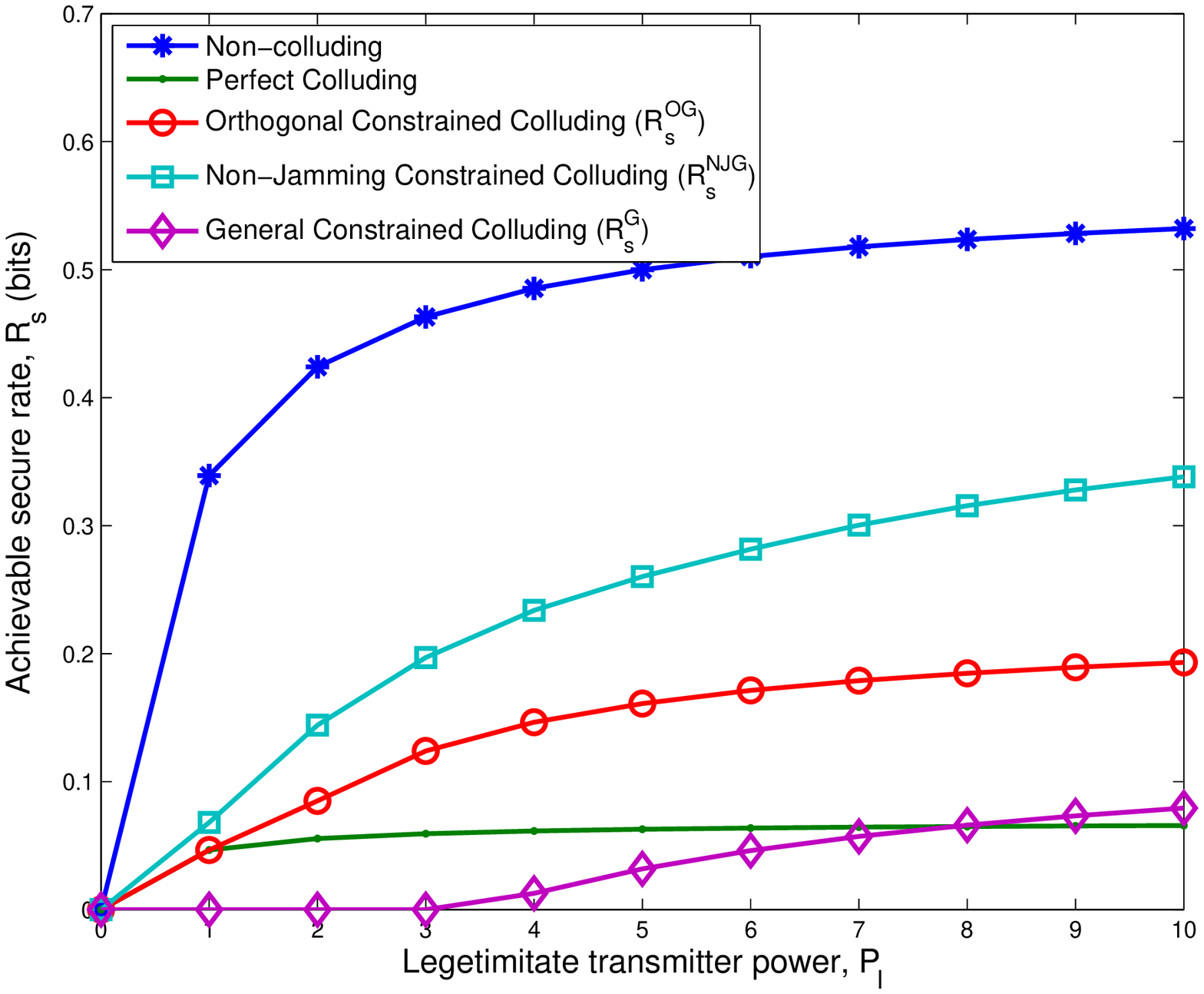}}\\[-5pt]
    \caption{Achievable secure rates $\Rc_s$ for $P_{je}=1$, $h_{je}^l=\sqrt{0.2}$, $h_{l}^{je}=h_{jm}=1$, $N_l=N_{je}=N_{je}^m=N_{je}^c=1,j\in\{1,2\}$.}%
    \label{fig:Gaus_Case1}
\end{figure*}
Fig.~\ref{fig:Gaus_Case1} compares the secure rates for the Gaussian WTC-CCE, i.e., $\Rc_s^{G},\Rc_s^{OG},\Rc_s^{NJG}$, to the non-colluding and perfect colluding scenarios in two different collusion channel conditions. It can be seen that the perfect colluding assumption significantly overestimates the eavesdroppers. Recall that the WTC-CCE rates consider the best possible strategy for the eavesdroppers; which may not be achievable for them. Thus, even the constrained colluding rates consider the worst case scenarios for the eavesdroppers' ability to collude.

In Fig.~\ref{fig:Gaus_Case1A} (weak collusion channel), using the orthogonal collusion channel for eavesdroppers is worse than using the non-orthogonal one (because $\Rc_s^{OG}\geq \Rc_s^{NJG}$). In fact, with weak direct collusion links, eavesdroppers may benefit of the main channel by relaying (transmitting correlated codewords). Hence, the optimal $\rho_1,\rho_2$ for $\Rc_s^{NJG}$ are not zero; while they are zero for $\Rc_s^{OG}$. However, for improved collusion channel (in Fig.~\ref{fig:Gaus_Case1B}), using an orthogonal collusion channel is better (from the eavesdroppers point of view) if one cannot use jamming (or does not want to use jamming to avoid exposure), i.e., $\Rc_s^{OG}\leq \Rc_s^{NJG}$. To evaluate the general rate $\Rc_s^{G}$, one should note the effect of jamming in addition to collusion, which even enables the eavesdroppers (or now jammers) to make the secure rate zero for some range of legitimate power $P_l$.


\section{Conclusion}\label{sec:discussion}
We proposed WTC-CCE, a wiretap-based channel model to capture collusion constraints and derived the achievable secure rates. Our results showed that indeed the perfect collusion model overestimates the eavesdroppers if they choose to be unexposed. With no exposure constraint, they can jam to further reduce the secure rate in some cases.

\end{document}